\documentclass{aa}  

\usepackage{graphicx}
\usepackage{txfonts}
\usepackage{hyperref}
\hypersetup{colorlinks=true,urlcolor=black, citecolor=blue, linkcolor=black}

\usepackage{multirow}
\usepackage{todonotes}
\usepackage{soul}
\usepackage{comment}
\usepackage{orcidlink}
\usepackage{subcaption,booktabs}

\usepackage{natbib}

\bibpunct{(}{)}{;}{a}{}{,} 

\definecolor{steelblue}{rgb}{0.274 0.510 0.706}

\defcitealias{Sabhahit2023}{S23}
\defcitealias{Vink2001}{V01}
\defcitealias{Briel2022}{B22}
\defcitealias{Tanikawa2023}{T23}
\defcitealias{Gabrielli2024}{G24}
\defcitealias{Chen2015}{C15}
\begin{document} 

   \title{Impact of stellar winds on the pair-instability supernova rate}

    \titlerunning{Pair-instability supernova rate}

   \subtitle{}

   \author{Filippo Simonato \inst{1,2,3}\thanks{e-mail: \href{mailto:filippo.simonato@gssi.it}{filippo.simonato@gssi.it}}
   \and
   Stefano Torniamenti \inst{4,5}\thanks{e-mail: \href{mailto:sttorniamenti@mpia.de}{sttorniamenti@mpia.de}}
   \and
   Michela Mapelli\thanks{e-mail: \href{mailto:mapelli@uni-heidelberg.de}{mapelli@uni-heidelberg.de}}\inst{1,2,5,6,7}\orcidlink{0000-0001-8799-2548}
   \and
   Giuliano Iorio\inst{8}\orcidlink{0000-0003-0293-503X}
   \and 
   Lumen Boco\inst{5}
   \and
   Franca De Domenico-Langer\inst{5}
   \and
   Cecilia Sgalletta\inst{9,10,11}\orcidlink{0009-0003-7951-4820}
          }\authorrunning{F. Simonato et al.}
    
    \institute{
    $^{1}$Physics and Astronomy Department Galileo Galilei, University of Padova, Vicolo dell'Osservatorio 3, I--35122, Padova, Italy\\
    $^{2}$Gran Sasso Science Institute (GSSI), Viale Francesco Crispi 7, 67100 L’Aquila, Italy\\
    $^{3}$INFN, Laboratori Nazionali del Gran Sasso, 67100 Assergi, Italy\\
    $^{4}$Max-Planck-Institut f{\"u}r Astronomie, K{\"o}nigstuhl 17, 69117, Heidelberg, Germany\\
    $^{5}$Universit\"at Heidelberg, Zentrum f\"ur Astronomie (ZAH), Institut f\"ur Theoretische Astrophysik, Albert-Ueberle-Str. 2, 69120, Heidelberg, Germany\\
    $^{6}$ Universit\"at Heidelberg, Interdisziplin\"ares Zentrum f\"ur Wissenschaftliches Rechnen, Heidelberg, Germany\\
    $^7$INFN, Sezione di Padova, Via Marzolo 8, I--35131 Padova, Italy\\
    $^8$Departament de Física Quàntica i Astrofísica, Institut de Ciències del Cosmos, Universitat de Barcelona, Martí i Franquès 1, E-08028 Barcelona, Spain\\
    $^9$SISSA, via Bonomea 365, I–34136 Trieste, Italy \\
    $^{10}$National Institute for Nuclear Physics – INFN, Sezione di Trieste, I--34127 Trieste, Italy \\
    $^{11}$Istituto Nazionale di Astrofisica – Osservatorio Astronomico di Roma, Via Frascati 33, I--00040, Monteporzio Catone, Italy
    }

   \date{Received XXXX; accepted YYYY}

 
  \abstract
{Very massive stars (VMSs, $M_{\star}$ $\geq$ 100 M$_{\odot}$) play a crucial role in several astrophysical processes. At low metallicity, they might 
   collapse directly into black holes, or end their lives as pair-instability supernovae. 
      Recent observational results set an upper limit of $0.7\,{}\mathrm{ yr}^{-1} \,{}\mathrm{ Gpc}^{-3}$ on the  rate density of pair-instability supernovae in the nearby Universe. However, most theoretical models predict rates  exceeding this limit. 
      Here, we compute new VMS tracks with the \textsc{mesa} code, and use them to analyze the evolution of the (pulsational) pair-instability supernova rate density across cosmic time. 
   We show that stellar wind models accounting for the transition between optically thin and  thick winds  yield a pair-instability supernova rate $\mathcal{R}_{\mathrm{PISN}}\sim{}0.1$ Gpc$^{-3}$ yr$^{-1}$ at redshift $z\sim{}0$, about two orders of magnitude lower than our previous models. We find that the main contribution to the pair-instability supernova rate comes from stars with metallicity $Z\sim{}0.001-0.002$. Stars with higher metallicities cannot enter the pair-instability supernova regime, even if their zero-age main sequence mass is up to 500~M$_\odot$. The main reason is that VMSs enter the regime for optically thick winds during the main sequence at metallicity as low as $Z\sim{4}\times{}10^{-4}$. This enhances the mass loss rate, quenching the growth of the He core and thus preventing the onset of pair-instability in later evolutionary stages. 
   This result highlights the critical role of mass loss  in shaping the final fate of very massive stars and the rate of pair-instability supernovae.
}

   \maketitle

\section{Introduction}
Very massive stars (VMSs), with an initial mass  $>100 \text{ M}_{\odot}$,  
represent a challenge to our understanding of stellar evolution due to their 
extreme physical properties. Several studies have explored their 
evolutionary paths \citep{Belkus2007,Yungelson2008,Grafener2011,Brott2011,Georgy2013,Yusof2013,Hirschi2015,Limongi2018} and their observational properties in the local Universe  
\citep{Crowther2010, Bestenlenher2014}. 

As VMSs evolve, they 
lose mass via stellar winds, including 
nitrogen and other products of hot hydrogen burning. Hence, they might be one of the main sources of the N enhancement observed in some compact high-redshift galaxies \citep{Vink2023,Cameron2023,Charbonnel2023,Gieles2025}. Several works show that, if they directly collapse onto a black hole, VMSs can form intermediate-mass black holes ({black holes}  with mass $>10^2$ M$_\odot$,  \citealt{Ebisuzaki2001,Heger2002,Portegies2002,Portegies2004,Portegies2005,Giersz2015,Mapelli2016,Spera2017,Costa2023}).

If VMSs do not undergo direct collapse at the end of their life but instead explode as supernovae, they release enormous amounts of material into the interstellar medium 
\citep{Ohkubo2006,Yokoyama2012,Kojima2021,Goswami2022,Vink2023}. 
Specifically, one of the possible final fates of VMSs is a pair-instability supernova (PISN) \citep{Heger2002, Woosley2017, Spera2017}. 
PISNe occur when 
the efficient formation of electron-positron pairs softens the equation of state of the carbon-oxygen core \citep{Barkat1967,Fraley1968,Bond1984,Heger2002}. 
The resulting contraction of the core  triggers explosive burning of oxygen and silicon. As a 
consequence, the star may undergo a single, powerful explosion that leaves no compact remnant (PISN), or it may experience multiple pulsations 
before ultimately collapsing into a black hole (pulsational pair-instability supernova, PPISN; \citealp{Woosley2002,Chen2014,Yoshida2016,Woosley2017}). Since the onset of the instability depends on the central density and temperature, the final mass of the helium core ($M_{\mathrm{He,f}}$) is a  good proxy to infer whether a star enters the pair-instability regime or not: according to \cite{Woosley2017}, if $32 \text{ M}_{\odot} \leq M_{\mathrm{He,f}} < 64 \text{ M}_{\odot}$, it undergoes a PPISN; if $64 \text{ M}_{\odot} \leq M_{\mathrm{He,f}} < 135 \text{ M}_{\odot}$, it explodes in a PISN; otherwise, if $M_{\mathrm{He,f}} > 135 \text{ M}_{\odot}$, it collapses directly onto an intermediate-mass BH. 

PISNe and PPISNe may have a deep impact on the mass function of black holes, as they are expected to open a gap between $\approx{60}$ and $\approx{120}$ M$_\odot$ \citep{Woosley2017,Spera2017}. Our knowledge of these boundaries is however affected by multiple sources of uncertainties, concerning nuclear reaction rates, stellar rotation, convection, and the energetics of failed supernovae \citep[e.g.,][]{Leung2019,Farmer2019, Farmer2020, Mapelli2020,Marchant2020, Renzo2020, Song2020, Tanikawa2021, Farrell2021, Vink2021, Woosley2021, Farag2022, Hendriks2023,Winch2025}. 

As for observations, we still lack an uncontroversial detection of a (P)PISN, even if several studies reported possible  candidates  \citep{Woosley2007, Gal-Yam2009, Quimby2011, Cooke2012,Kozyreva2015,Kozyreva2018,Mazzali2019}. 
The supernova SN2018ibb is possibly the strongest PISN candidate reported to date \citep{Schulze2024}. The same authors estimated that the rate density of PISNe $\mathcal{R}_{\mathrm{PISN}}$ should lie between $0.009 \text{ yr}^{-1} \text{ Gpc}^{-3}$  and $0.7 \text{ yr}^{-1} \text{ Gpc}^{-3}$ at 2$\sigma{}$ confidence, based on the spectroscopically complete ZTF Bright Transient survey of superluminous supernovae \citep{Fremling2020, Perley2020}.  

Theoretical models estimate the PISN rate to be between $10^{-5}$ and $10^{-2}$ times the core-collapse supernova rate \citep{Langer2007,Briel2022}, but such estimates are affected by uncertainties about the initial mass function and metallicity of the progenitors \citep{Langer2007},  the cosmic star formation history \citep{Briel2022,Gabrielli2024}, 
 models of stellar evolution \citep{Takahashi2018,duBuisson2020,Farmer2020,Tanikawa2023,Gabrielli2024}, and whether the progenitor evolves as an isolated
star or in a binary system \citep{Briel2022,Tanikawa2023}.

\citet[][hereafter  \citetalias{Briel2022}]{Briel2022} investigated how star-forming environments influence electromagnetic and gravitational-wave transients by employing empirical models \citep{Madau2014} and cosmological simulations (\textsc{EAGLE}, \citealp{Schaye2015,Crain2015}; \textsc{IllustrisTNG}, \citealp{Springel2005,Nelson2018,Pillepich2018,Naiman2018,Marinacci2018}; \textsc{Millennium}, \citealp{Springel2005}). 
Among their models, the only one consistent with the upper limit found by \cite{Schulze2024} 
is based on the  \textsc{Eagle} simulation \citep{Schaye2015}, combined with the \textsc{B-Pass} population synthesis models \citep{Eldridge2017}. In a follow-up study, \cite{Briel2024}  focus on the \textsc{IllustrisTNG} simulation and obtain an expected PISN rate density  $\mathcal{R}_{\mathrm{PISN}}=2-29$ Gpc$^{-3}$ yr$^{-1}$ at redshift $z=0$, by quantifying the uncertainties from their binary evolutionary models.

\citet[][hereafter  \citetalias{Tanikawa2023}]{Tanikawa2023} explored the influence of the $^{12}$C($\alpha$,$\gamma$)$^{16}$O reaction rate on the PISN rate in models of binary populations. By adopting the standard reaction rate and assuming a maximum zero-age main-sequence (ZAMS) mass of 300 M$_{\odot}$, they found $\mathcal{R}_{\mathrm{PISN}} \geq 1 \mathrm{\,yr}^{-1} \mathrm{ Gpc}^{-3}$. In contrast, using a $3\sigma$ lower reaction rate, they found $\mathcal{R}_{\mathrm{PISN}} \ll 1 \mathrm{\,yr}^{-1} \mathrm{ Gpc}^{-3}$. The critical role of the $^{12}$C($\alpha$,$\gamma$)$^{16}$O reaction rate has been largely debated in the context of the pair-instability mass gap of black holes, with lower rates leading to higher black hole masses below the gap  \citep[e.g.,][]{Farmer2019,Farmer2020,Costa2021,Woosley2021}. 

\citet[][hereafter  \citetalias{Gabrielli2024}]{Gabrielli2024} 
explored various criteria for the onset of pair instability, including
the impact of the   metallicity threshold 
for PISNe, the maximum ZAMS mass, and the dispersion in the metallicity distribution of the host galaxies. 
They found that all these quantities  affect the PISN rate, resulting in a seven orders of magnitude variation up to $\mathcal{R}_{\mathrm{PISN}} \sim 2000 \text{ yr}^{-1} \mathrm{ Gpc}^{-3}$. According to their analysis (Fig.~8 from \citealt{Gabrielli2024}), only models with a metallicity threshold  for  the onset of pair instability  $Z\leq{}1.5\times{}10^{-3}$, a maximum ZAMS star mass $M_\textrm{ZAMS}\leq{}150$ M$_\odot$, and a narrow metallicity spread yield rates $\mathcal{R}_{\mathrm{PISN}}<1$ Gpc$^{-3}$ yr$^{-1}$.
This result confirms that the PISN rate is very sensitive to the assumed metallicity threshold for the onset of PISNe. Moreover, assuming a narrow metallicity dispersion suppresses pockets of low-metallicity star formation in the nearby Universe \citep{Boco2019,Santoliquido2022}. 

In summary, most theoretical estimates of the PISN rate density 
exceed the upper limit reported by \cite{Schulze2024}. The only models that yield PISN rates below this upper limit assume low values for the $^{12}$C($\alpha$,$\gamma$)$^{16}$O reaction rate,  enforce a low metallicity threshold for the onset of PISNe, 
 suppress the low-metallicity tails of star formation in the nearby Universe, and/or put a cap on the maximum ZAMS mass. The upper limit to the maximum ZAMS mass is particularly critical, because we have observed stars with $M_{\textsc{ZAMS}}>150$ M$_\odot$ even in the local Universe \citep{Crowther2010,Bestenlenher2014,Bestenlehner2020b}.


Here, we explore an alternative scenario inspired by recent models of optically thick winds for VMSs \citep[][hereafter \citetalias{Sabhahit2023}]{Sabhahit2023}. This scenario does not require to fine tune the $^{12}$C($\alpha$,$\gamma$)$^{16}$O reaction rate, works for a realistic evolution of the metallicity spread in galaxies, and does not require to cap the maximum star mass, while it naturally leads to a low metallicity threshold $Z_{\mathrm{th}}$ for the activation of PISNe ($Z_{\mathrm{th}} \sim{Z_{\odot}/10}$, \citetalias{Sabhahit2023}).

{
Although the stellar winds of massive and very massive stars remain an active topic of research \citep{Yang2023, Cheng2024, Pauli2025},} recent models of mass loss by stellar winds \citep{Vink2011,Sabhahit2022,Sabhahit2023} indicate that VMSs develop optically thick winds already on the main sequence, becoming WNh stars \citep{Bestenlehner2020b}. The transition between optically thin winds of O-type stars and optically thick ones is a function of the electron-scattering Eddington ratio $\Gamma_{\mathrm{Edd}}$. At the metallicity of the Large Magellanic Cloud and even the Small Magellanic Cloud, this model strips the H-rich envelope of VMSs completely, without the need for mass loss through binary interaction. Here, we investigate the impact of such new stellar wind models  (\citetalias{Sabhahit2023}) on the expected PISN and PPISN rate density. 
We show that a larger mass-loss rate for VMSs at intermediate metallicity (down to $Z\sim{}0.002$, \citetalias{Sabhahit2023}) seems to be the key to predicting PISN rates in agreement with the observational constraints.

This paper is organized as follows. We briefly describe the physics and the prescription adopted to the (P)PISN rate density in Sec. \ref{sec:method}. In Sec. \ref{sec:results}, we show the rate densities obtained with the two different stellar models. We discuss the impact of an enhanced mass-loss rate in Sec. \ref{sec:discussion}. Finally, our results are summarized in Sec. \ref{sec:summary}.






\section{Method}\label{sec:method}

\begin{figure}[h!]
    \centering
    \includegraphics[width=.5\textwidth]{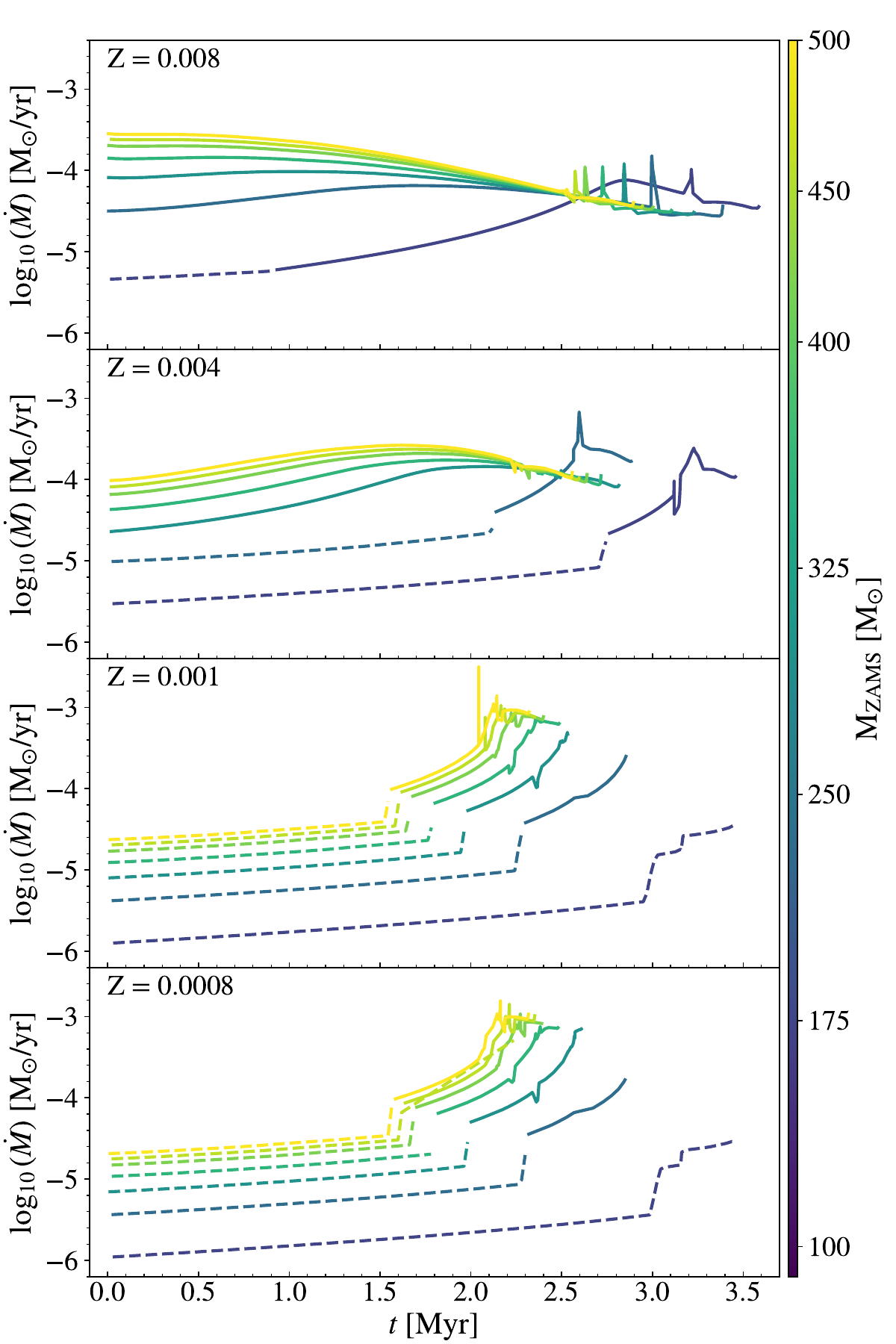}
    \caption{Mass-loss rate of our \textsc{mesa} models as a function of time, from the ZAMS until the end of core He burning. The models adopt the winds by \protect\cite{Sabhahit2023}. From top to bottom: $Z = 0.008$, 0.004, 0.001, and 0.0008. The solid lines correspond to the evolutionary phase in which the stellar models are in the optically thick regime ($\protect\Gamma_{\protect\mathrm{Edd}} > \protect\Gamma_{\protect\mathrm{Edd, tr}}$), while the dashed lines mark the optically thin regime.}
    \label{fig:gamma}
\end{figure}

We use the stellar evolution code \textsc{mesa} (version r12115; \citealp{Paxton2011, Paxton2013, Paxton2015, Paxton2018, Paxton2019, Jermyn2023}) to model massive stars and VMSs and produce stellar tables for the population synthesis code \textsc{sevn}  \citep{Spera2017, Spera2019, Mapelli2020, Iorio2023}. Then, we combine our population synthesis models with a metal-dependent star formation rate density history by means of the semi-analytic code \textsc{cosmo$\mathcal{R}$ate} \citep{Santoliquido2021, Santoliquido2022}. Here below, we describe our methodology in detail.

\subsection{Stellar models with \textsc{mesa}}\label{sec:mesa}

We adopt the \textsc{mesa} mass-loss scheme and parameter setup introduced by  \citetalias{Sabhahit2023}\footnote{We refer to the \textsc{mesa} inlists and run\_star\_extras files by \citetalias{Sabhahit2023}, which are publicly available at \href{{https://github.com/Apophis-1/VMS_Paper2}}{this gitlab repository}.}, with some changes to enforce  convergence until the formation of the CO core. In the following, we give an overview of our main parameters and initial conditions.

During the main sequence, we use the mass-loss framework described by \citet{Sabhahit2022} and extended to metal-poor stars by \citetalias{Sabhahit2023}. This model describes the transition from an optically thin 
to an optically thick 
wind in VMSs \citep{Vink2011,Bestenlenher2014}, exploiting the concept of transition mass-loss rate \citep{Vink2012}. 
The transition happens when $\Gamma_{\mathrm{Edd}} \geq \Gamma_{\mathrm{Edd, tr}}$, where $\Gamma_{\mathrm{Edd}}$ is the Eddington ratio of a star of luminosity $L_\ast$, defined as the ratio between $L_\ast$ and the Eddington luminosity of the star, whereas $\Gamma_{\mathrm{Edd, tr}}$ is the value of the Eddington ratio of the same star  computed at the transition point. When stellar winds become optically thick, the mass-loss rate is enhanced according to \cite{Vink2011}:
\begin{equation}\label{eq:mloss}
    \dot{M} = \dot{M}_{\mathrm{tr}}\,\left(\frac{L}{L_{\mathrm{tr}}}\right)^{4.77}\,\left(\frac{M}{M_{\mathrm{tr}}}\right)^{-3.99}\,,
\end{equation}
{where $\dot{M}_{\mathrm{tr}}$, $L_{\mathrm{tr}}$, and $M_{\mathrm{tr}}$ are the mass-loss rate, the luminosity, and the mass at the transition point.} To obtain $\Gamma_{\mathrm{Edd, tr}}$, the following condition is iteratively verified:
\begin{equation}
    \eta_{\mathrm{Vink}} \equiv \frac{\dot{M}_{\mathrm{Vink}}\,v_{\infty}}{L / c}\, = \frac{\eta}{\tau_{F,s}}\left(\frac{v_{\infty}}{v_{\rm{esc}}}\right) \sim \left(1 + \frac{v_{\rm{esc}}^2}{v_{\infty}^2}\right)^{-1}\,,
\end{equation}
where $\dot{M}_{\mathrm{Vink}}$ is the mass-loss rate computed following \cite{Vink2001}, $v_{\infty}$ the asymptotical velocity, $v_{\mathrm{esc}}$ the escape velocity, $\eta$ the wind efficiency of the model, and $\tau_{\mathrm{F}}$ the flux-weighted mean optical depth. This condition allows to evaluate $L_{\mathrm{tr}}$ at each iteration, necessary to compute $\Gamma_{\mathrm{Edd, tr}}$.

During the core He-burning phase, we use Eq.~\ref{eq:mloss} only for temperatures between $4\times10^3 \mathrm{ K}$ and {
$10^5$ K}, while for lower temperatures ($T<4\times10^3 \mathrm{ K}$), we follow \citet{deJager1988}. {
For $T>10^5$ K and hydrogen mass fraction $X_{\text{C}}<0.01$}, we adopt the Wolf-Rayet (WR)  mass-loss models by \citet{Sander2020}. {Recently, \citet{Boco2025} showed that this formalism, when combined with fast rotation, can strip single metal-poor O-type stars and convert them into WNh stars even at low  metallicity ($Z\sim{0.0002}$). Thus, this mechanism offers  a potential explanation for the presence of WR stars in metal-poor galaxies, such as the Small Magellanic Cloud \citep{Schootemeijer2021}.}

We treat convective mixing with the mixing length theory (MLT) by \citet{Cox1968}, with a constant mixing-length ratio $\alpha_{\mathrm{MLT}} = 1.5$. The convective boundaries are set by applying the Ledoux criterion \citep{Ledoux1947}. 
We use the semi-convective diffusion parameter $\alpha_{\mathrm{SC}} = 1$.  We choose the exponential overshooting by \citet{Herwig2000} to describe the overshooting above the convective regions with efficiency parameter $f_{\mathrm{OV}} = 0.03$. We consider only non-rotating models. Finally, we turn on the MLT++ option \citep{Paxton2013}  only during the core He-burning phase to ensure convergence.  MLT++ is an option that allows to reduce superadiabaticity ($\nabla_{\mathrm{T}} > \nabla_{\mathrm{ad}}$, where $\nabla_{\mathrm{T}}$ is the temperature gradient and $\nabla_{\mathrm{ad}}$ the adiabatic temperature gradient) in radiation-dominated convective regions with a-posteriori calculations. We refer to \citetalias{Sabhahit2023} for a more in-depth description of the parameters and the model.

We make exactly the same assumptions as \citetalias{Sabhahit2023} with only two exceptions: the definition of core boundaries and the nuclear reaction network. As for the core boundaries, we define the radius of the helium core as the radius inside which the mass fraction of hydrogen drops below $f_{\rm X}=10^{-4}$. Similarly, we define the radius of the helium (carbon) core as the radius inside which the mass fraction of $^{4}$He ($^{12}$C) drops below   $f_{\rm Y}=10^{-4}$ ( $f_{\rm C}=10^{-4}$). The default of \textsc{mesa} instead sets the boundary of the H, He, and C core where the mass fraction of H, $^{4}$He, and $^{12}$C  goes  below 0.1, respectively. Our definition yields a more precise description of the core and  is the same as assumed by the \textsc{MIST} libraries \citep{Paxton2011,Paxton2013,Paxton2015,Dotter2016,Choi2016}.  

We adopt the default nuclear reaction network in \textsc{mesa} with one exception:  after the main sequence, we enable the \verb|co_burn| net (parameter \verb|auto_extend_net = true|), which includes $^{28}$Si, in order to have a more complete and extended network for C- and O-burning and $\alpha$-chains with respect to \citetalias{Sabhahit2023}.
    

We calculate stellar models with  initial masses ranging from 50  to 500 M$_{\odot}$ by intervals of 25 M$_{\odot}$. The initial metallicities are $Z$ = 0.008, 0.006, 0.004, 0.002, 0.001, 0.0008, 0.0006, 0.0004, 0.0002, and 0.0001. Thus, our grid consists of 190 models (19  initial masses and 10 initial metallicities).
The initial helium mass fraction, $Y$, is computed as $Y = Y_{\mathrm{prim}} + (\Delta Y / \Delta Z),Z$, where $Y_{\mathrm{prim}}$ is the primordial helium abundance and $\Delta Y / \Delta Z$ describes the helium-to-metal ratio. We adopt the default values from \textsc{mesa} \citep{Pols1998}, with $Y_{\mathrm{prim}} = 0.24$ and $\Delta Y / \Delta Z = 2$. This formulation spans the range from a nearly primordial composition ($X = 0.76$, $Y = 0.24$, $Z = 0$) to a quasi-solar one ($X = 0.70$, $Y = 0.28$, $Z = 0.02$). The hydrogen mass fraction is given by $X = 1 - Y - Z$.

We evolve all the stellar models until the luminosity due to carbon burning $L_\mathrm{C}$ is less than the 20\% of the total luminosity $L_{\mathrm{TOT}}$ ($L_\mathrm{C}/L_{\mathrm{TOT}} < 0.2$) and the central He abundance is $Y_\mathrm{c}<10^{-8}$. 

Figure~\ref{fig:gamma} shows the mass-loss rate of our models for four selected metallicities, until the end of core He burning. During the main sequence, the initial mass required to enter the optically thick wind regime grows with decreasing metallicity: at $Z = 0.0008$, models with $M_{\mathrm{ZAMS}} \leq 125 \text{ M}_{\odot}$ do not experience optically thick winds, while at $Z = 0.008$ even the less massive models experience enhanced winds. 

When stars that begin their life with $\Gamma_{\mathrm{Edd}} < \Gamma_{\mathrm{Edd, tr}}$ enter the optically thick wind regime, they experience a boost in the mass-loss. Instead, the mass-loss rate of stars that begin their life with $\Gamma_{\mathrm{Edd}} > \Gamma_{\mathrm{Edd, tr}}$ reaches a peak of $\sim 10^{-3.5} \text{  M}_{\odot}\mathrm{ yr}^{-1}$ during the main sequence and then starts to decrease. We discuss other properties of our \textsc{mesa} stellar models in Appendix \ref{app}.

\subsection{Population synthesis with \textsc{sevn}}\label{sec:sevn}

We use the population-synthesis code \textsc{sevn} (\citealp{Spera2017, Spera2019, Mapelli2020,Iorio2023})\footnote{
We used the version 2.10.1 of \textsc{sevn}, updated to commit hash c9dc8e4578990fd85cba91e082ebff6068388f56. \textsc{sevn} is publicly available at \href{https://gitlab.com/sevncodes/sevn}{{this gitlab repository}}.} to 
interpolate among our \textsc{mesa} models and estimate the (P)PISN rate density. \textsc{sevn} 
evolves populations of single and binary stars by interpolating pre-computed stellar tracks and incorporating semi-analytic models for mass transfer, tides, supernova explosions, and gravitational-wave decay. 
We refer to \citet{Iorio2023} for a complete description of the code.


Here, we use the outputs from 
our new \textsc{mesa} models 
to obtain new 
evolutionary tables for stars with mass $M_{\star} \geq 50 \mathrm{ M}_{\odot}$. 

We consider the same initial metallicities as in the \textsc{mesa} models. We simulate $10^5$ stars for each metallicity. The initial masses are distributed following a Kroupa initial mass function (IMF, \citealp{Kroupa2001}) with slope $\alpha = 2.3$ between 50 M$_{\odot}$ and 500 M$_{\odot}$. 
To model pair-instability and pulsational pair-instability supernovae, 
we use the fitting formulas described by 
\cite{Mapelli2020}, based on the models by \citet{Woosley2017}. Namely, a star undergoes a PPISN if the final mass of the He core is between $32 \text{ M}_{\odot} \leq M_{\mathrm{He,f}} < 64 \text{ M}_{\odot}$. A PISN takes place if $64 \text{ M}_{\odot} \leq M_{\mathrm{He,f}} \leq 135 \text{ M}_{\odot}$. For higher masses of the He core, the star directly collapses onto an intermediate-mass BH. 

As a comparison, we also consider the \textsc{sevn} tables produced with the stellar evolution code \textsc{parsec} \citep{Bressan2012,Chen2015, Costa2019b, Costa2019a, Nguyen2022,Costa2025}, and described by \cite{Iorio2023}. The main difference between the \textsc{parsec} and \textsc{mesa} models is that the former adopt a different stellar wind model from the one 
described by \citetalias{Sabhahit2023}. 
The \textsc{parsec} wind model introduced by \citet[][hereafter  \citetalias{Chen2015}]{Chen2015} is based on the fitting formulas by \cite{Vink2000} and \cite{Vink2001} for the winds of O-type stars, with additional corrections for electron scattering, and a dependence of the mass-loss on the Eddington ratio \citep{Graefener2008,Vink2011}.
We provide more details about the \textsc{parsec} models in Appendix~\ref{app:parsec}.

\subsection{(P)PISN rate density evolution across cosmic time}

We model the evolution of the (P)PISN rate density as a function of  redshift with the code \textsc{cosmo$\mathcal{R}$ate}\footnote{\textsc{cosmo$\mathcal{R}$ate} is publicly available at \href{https://gitlab.com/Filippo.santoliquido/cosmo_rate_public}{{this gitlab repository}}.} \citep{Santoliquido2020, Santoliquido2021}, 
based on the observed metal-dependent cosmic star formation rate density \citep{Madau2017}.

The default version of \textsc{cosmo$\mathcal{R}$ate} 
estimates the merger rate density of compact binaries as a function of the redshift, but it can be adapted to trace the rate density of other transients of stellar origin, including (P)PISNe. 
Specifically, we calculate the rate of (P)PISNe as follows: 
\begin{equation}
\label{eq:mrd}
    \mathcal{R}(z) = \int_{z_{{\rm{max}}}}^{z}\left[\int_{Z_{{\rm{min}}}}^{Z_{{\rm{max}}}} \,{}\mathcal{S}(z',Z)\,{} 
    \mathcal{F}(z',z,Z) \,{}{\rm{d}}Z\right]\,{} \frac{{{\rm d}t(z')}}{{\rm{d}}z'}\,{}{\rm{d}}z',
\end{equation}
where $\mathcal{S}(z',Z)= \psi(z')\,{}{p(Z|z')}$. 

Here, $\psi(z')$ is the cosmic  star formation rate density at redshift $z'$, and {$p(Z|z')$} is the distribution of metallicity $Z$ at fixed formation redshift $z'$. In Equation \ref{eq:mrd}, ${{\rm{d}}t(z')}/{{\rm{d}}z'} =  H_{0}^{-1}\,{}(1+z')^{-1}\,{}[(1+z')^3\Omega_{\mathrm{M}}+ \Omega_\Lambda{}]^{-1/2}$, where $H_0$ is the Hubble  parameter, $\Omega_\mathrm{M}$ and $\Omega_\Lambda$ are the matter and energy density, respectively. We adopt the values from \cite{Planck2018}.

The term $\mathcal{F}(z',z, Z)$ in Equation~\ref{eq:mrd} is given by:
\begin{equation}
\mathcal{F}(z',z, Z) = \frac{1}{M_{\ast{}}(Z)}\,{}\frac{{\rm{d}}\mathcal{N}(z',z, Z)}{{\rm{d}}t(z)},
\end{equation}
where  ${{\rm{d}}\mathcal{N}(z',z, Z)/{\rm{d}}}t(z)$ is the rate of (P)PISNe from stars with metallicity $Z$ that formed at redshift $z'$  and undergo a (P)PISN at redshift $z$, and $M_\ast(z',Z)$ is the initial total stellar mass that forms at redshift $z'$ with metallicity $Z$. We obtain ${{\rm{d}}\mathcal{N}(z',z, Z)/{\rm{d}}}t(z)$  directly from our simulations. {Note that $\mathcal{N}(z',z, Z)$ also depends on the stellar lifetime (hence on the stellar mass) and on the assumed initial mass function. While we do not explicitly write down these dependences for notation simplicity, our catalogs fully account for them.}

We model the star formation rate density through the fitting formula by \citet{Harikane2022}:
\begin{eqnarray}
    \psi(z) = \left[61.7(1-z)^{-3.13}+10^{0.22(1+z)}+2.4^{0.5(1+z)-3}\right]^{-1}\frac{\text{M$_{\odot}$}}{\text{Mpc}^{3}\,\text{yr}}.
    \label{eq:sfrd}
\end{eqnarray}

They obtained this formula through the analysis of new measurements of the rest-frame UV luminosity at $z \sim 4-7$ based on wide and deep optical images obtained in the Hyper Supreme Cam (HSC) Subaru Strategic Program (SSP) survey \citep{Aihara2018} and the CFHT Large Area U-band Deep Survey (CLAUDS; \citealp{Sawicki2019}). This formula is a better match to recent James Webb Space Telescope (JWST) data \citep{Donnan2023,Harikane2023,Harikane2024} than the traditionally adopted fit by \citet{Madau2014}. 
 The star formation rate density is then multiplied by a correction factor $f_{\mathrm{Kroupa}} = 0.66$ to convert the SFR from a Salpeter IMF to a Kroupa IMF \citep{Madau2014}. 
 
    \label{eq:F}

The metallicity-evolution term {$p(Z|z')$} is the probability density function that a star formed at redshift $z'$ has a metallicity $Z$, modelled as a log-normal distribution function \citep{Chruslinska2019,Santoliquido2023}:
\begin{equation}
    {p(Z|z')} = \frac{1}{\sqrt{2\,\pi\,\sigma_\mathrm{Z}^2}}\,\exp{\left\{-\,\frac{\left[\log(Z/Z_{\odot}) - \langle{}\log(Z/Z_{\odot})\rangle{}\right]^2}{2\,\sigma_\mathrm{Z}^2}\right\}}\,.
    \label{eq:log_Z}
\end{equation}
We derive  $\langle{}\log(Z/Z_{\odot})\rangle{}$ from the fitting formula by \citet{Madau2017}
\begin{equation}
    \log\langle Z/Z_{\odot} \rangle = 0.153 - 0.074\,z^{1.34}\,,
    \label{eq:Z}
\end{equation} 
knowing that $\langle{}\log{Z/Z_\odot}\rangle{}=\log{\langle{}Z/Z_\odot\rangle{}}-\ln{(10)}\,{}\sigma_{\rm Z}^2/2$. We assume $\sigma_Z=0.4$, based on the results by \citet{Sgalletta2024}.





\begin{figure}[h!]
    \centering
    \includegraphics[width=.5\textwidth]{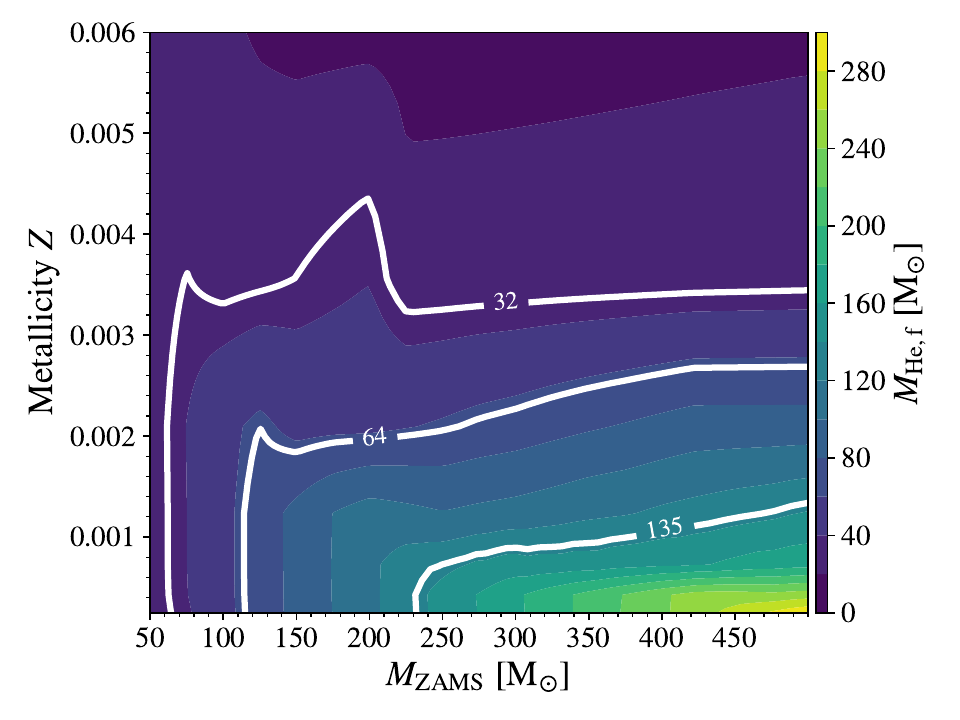}
    \caption{Contour plot showing the He core masses at the end of the He burning for our \textsc{mesa} tracks adopting \protect\citetalias{Sabhahit2023} winds  as a function of the metallicity and ZAMS mass. The white lines highlight the levels at $32 \protect\text{M}_{\odot}$, $ 64 \protect\text{ M}_{\odot}$ and $135 \protect\text{ M}_{\odot}$, which set the boundaries to have PPISNe, PISNe, and direct collapse via photodisintegration.}
    \label{fig:He_mass_cont}
\end{figure}

\begin{figure}[h!]
    \centering
    \includegraphics[width=.5\textwidth]{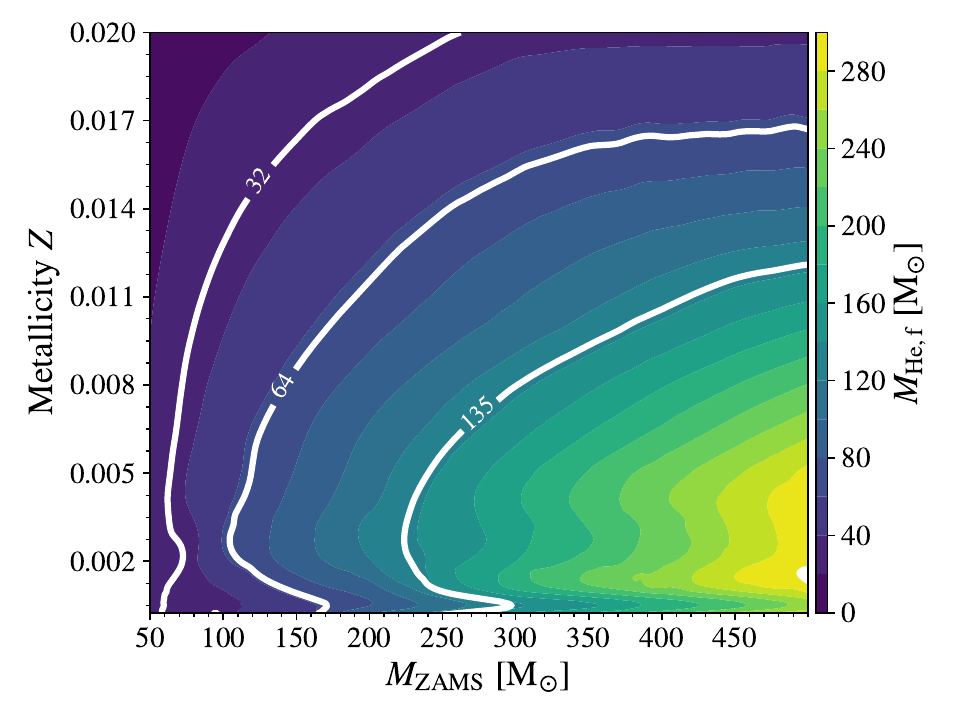}
    \caption{Same as Fig. \ref{fig:He_mass_cont}, but for models  \protect\citetalias{Chen2015}.}
    \label{fig:He_mass_cont_old}
\end{figure}

\section{Results} \label{sec:results}

\subsection{The metallicity threshold for (P)PISNe}\label{sec:threshold}


\begin{figure*}[h!]
    \centering
    \includegraphics[width=.9\textwidth]{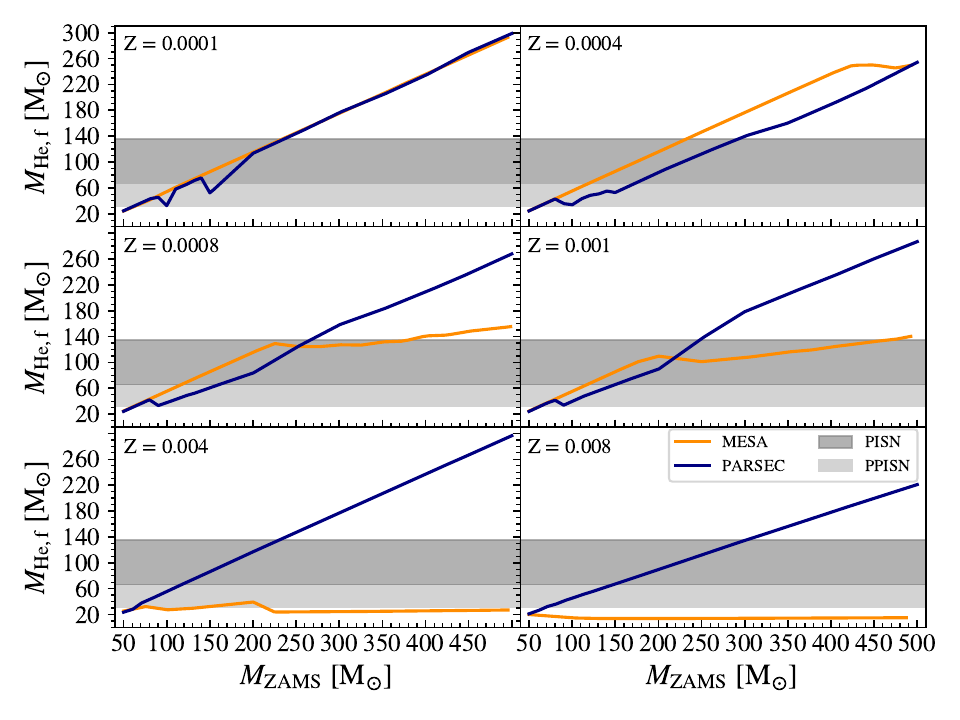}
    \caption{Final helium core mass as a function of the ZAMS  mass. The {orange} lines are our new \textsc{mesa} tracks (models \protect\citetalias{Sabhahit2023}), while the blue lines correspond to results from \textsc{parsec} tracks (models \protect\citetalias{Chen2015}). Dark (Light) gray shaded areas: regime for PISNe (PPISNe). From top left to bottom right we show metallicities $Z=0.0001$, 0.0004, 0.0008, 0.001, 0.004, 0.008.} 
    \label{fig:He_core_mesa_parsec}
\end{figure*}

Figure~\ref{fig:He_mass_cont} shows the final helium core mass for our \textsc{mesa} models with \citetalias{Sabhahit2023} winds as a function of metallicity and ZAMS mass. We highlight the levels corresponding to $M_\mathrm{He,f}=32,$ 64, and 135 M$_\odot$, because these correspond to the threshold for PPISN, PISN, and direct collapse according to \cite{Woosley2017}. In the \citetalias{Sabhahit2023} model, the maximum He core mass of stars with $M_\textrm{ZAMS}$ up to 500 M$_\odot$ is below 65 M$_\odot$ 
for metallicity $Z\gtrsim{}0.003$ and below 135 M$_\odot$ for $Z\gtrsim{}0.0015$. 

Assuming that the He core mass is a good proxy for the onset of pair instability and adopting the models by \cite{Woosley2017}, the results presented in  Figs.~\ref{fig:He_mass_cont} and ~\ref{fig:He_mass_cont_old} allow us to estimate the threshold metallicity for the onset of (P)PISNe for both the \citetalias{Sabhahit2023} and \citetalias{Chen2015} models. With our new models (\citetalias{Sabhahit2023}), the maximum metallicity for PISNe and PPISNe are $Z \sim{} 2 \times 10^{-3}$ and $Z \sim{ 4.5} \times 10^{-3}$, respectively. 
For comparison, 
the model adopting the \textsc{parsec} stellar tracks (\citetalias{Chen2015}) yields $Z \sim{} 1.7 \times 10^{-2}$ and $Z \gtrsim{}{2} \times 10^{-2}$ as thresholds for PISNe and PPISNe, respectively (Fig.~\ref{fig:He_mass_cont_old}). Such metallicities are extremely high, above the solar value ($Z\sim{0.015}$,  e.g., \citealt{Caffau2011}).

In the \citetalias{Sabhahit2023} model (Fig. \ref{fig:He_mass_cont}), even stars whose initial mass is higher than 250 M$_{\odot}$ explode as PISNe if their metallicity lies between 0.001 and 0.002. At lower metallicities, stars with ZAMS mass $>250$ M$_\odot$ collapse directly to black holes, because photodisintegration prevents the explosion, while  stars with ZAMS mass between 110 and 230 M$_{\odot}$  undergo PISNe. Stars with $64 \text{ M}_{\odot} \leq M_{\mathrm{ZAMS}} \leq 110 \text{ M}_{\odot}$ undergo PPISN for $Z \leq 0.004$ and this mass range extends up to $M_{\mathrm{ZAMS}} = 500 \text{ M}_{\odot}$ for $0.002 \leq Z \leq 0.004$.

Figure~\ref{fig:He_core_mesa_parsec} shows the evolution of the He core mass at the end of core He burning for some selected metallicities, comparing \citetalias{Sabhahit2023} and \citetalias{Chen2015}. From this Figure, it is apparent how the onset of optically thick winds during the main sequence progressively "erodes" the core of the most massive stars for $Z\geq{}0.0004$ in our \textsc{mesa} models. In contrast, the \textsc{parsec} models grow large cores even at relatively high metallicity ($Z\geq{}0.0004$) because of the different wind model.

However, at low metallicities ($Z\lesssim{}0.0004$), the \textsc{parsec} models tend to develop smaller cores than the \textsc{mesa} models for ZAMS masses $M_\mathrm{ZAMS}\geq{}80$ M$_\odot$. This difference arises from the different envelope undershooting and convection models adopted by \textsc{mesa} and \textsc{parsec}  (see Appendices~\ref{app} and \ref{app:parsec}), rather than by the wind model. Specifically,  the \textsc{parsec} tracks develop effective dredge-up episodes that exchange mass between the envelope and the He core, reducing the overall He core mass at the end of core He burning \citep{Costa2021}. 

\begin{figure}[!ht]
    \centering
    \includegraphics[width=0.5\textwidth]{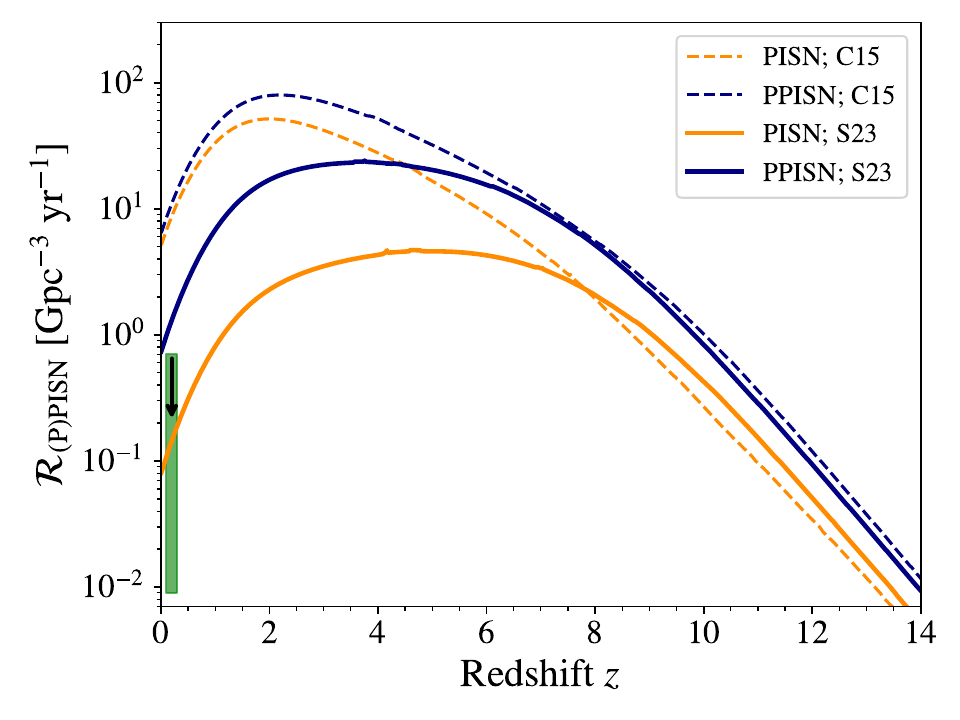}
    \caption{{PISN ({orange} lines)}  and PPISN (blue lines)  rate densities as a function of redshift. The solid and dashed lines are  obtained using the model by \protect\citetalias{Sabhahit2023} and \protect\citetalias{Chen2015}, respectively. The green shaded area represents the observational constraints set by \protect\citet{Schulze2024}. The black arrow highlights the upper limit from \protect\citet{Schulze2024}.}
    \label{fig:pisn_comp}
\end{figure}

\subsection{(P)PISN rate density}

\begin{figure*}
    \centering
    \includegraphics[width=\textwidth]{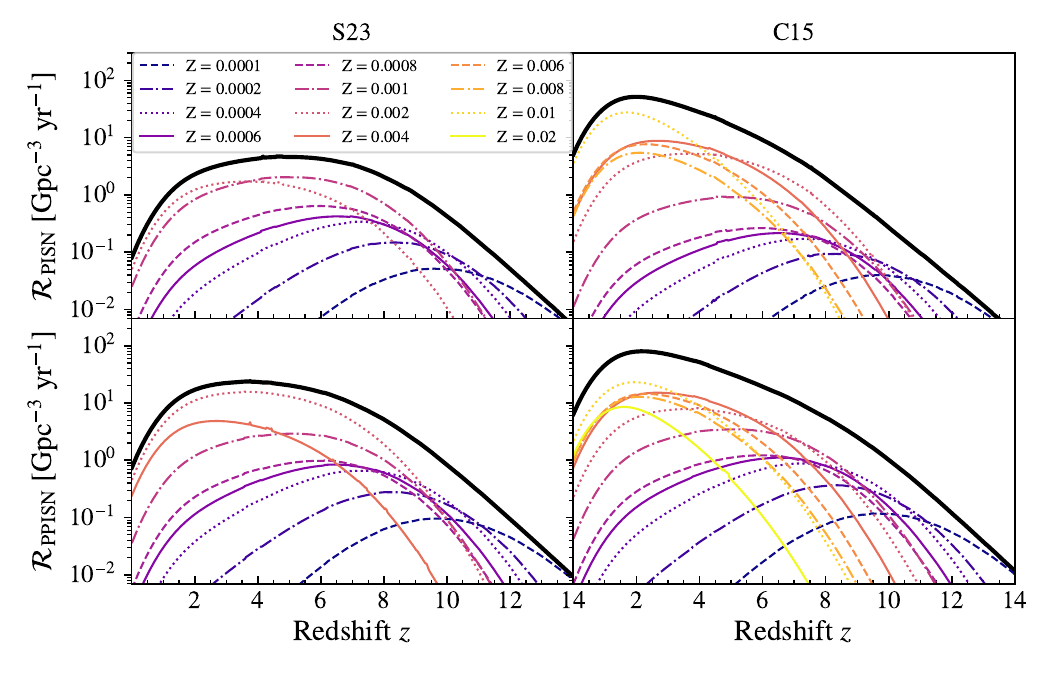}
    \caption{PISN (upper panels) and PPISN (lower panels) rate density evolution as a function of redshift.
    The left-hand and right-hand plots show the results obtained with the new wind model by \protect\citetalias{Sabhahit2023}  and the one by \protect\citetalias{Chen2015}, respectively. The black thick line represents the total rate density, whereas the colored lines show the contribution of individual metallicities from $Z = 1 \times 10^{-4}$ ({blue dashed} line) to $Z = 2 \times 10^{-2}$ ({solid} yellow line).} 
    \label{fig:pisn_new}
\end{figure*}

Figure \ref{fig:pisn_comp} shows the evolution of the PISN and PPISN rate density across cosmic time. We compare our new \textsc{mesa} models adopting the mass-loss rate by \citetalias{Sabhahit2023} to the \textsc{parsec} models with the mass-loss rate presented by \citetalias{Chen2015}. We also compare the models with the observational constraints reported by \citet{Schulze2024}. 

At low redshift, 
the (P)PISN rate density of the \citetalias{Sabhahit2023} model is lower by two orders of magnitude compared to the \citetalias{Chen2015} model (see Tab. \ref{tab4}). 
The PISN rate of the \citetalias{Sabhahit2023} model is below the observational upper limit set by \citet{Schulze2024}, whereas the \citetalias{Chen2015} model yields a much higher rate than the upper limit.

The main reason for this difference is that the metallicity threshold to activate the PISNe is lower in the \citetalias{Sabhahit2023} model ($Z\leq{}0.002$) than in the \citetalias{Chen2015} model ($Z\leq{}0.017$), as shown in Figs.~\ref{fig:He_mass_cont} and \ref{fig:He_mass_cont_old}. According to the \citetalias{Chen2015} model, we expect PISNe to take place even at the metallicity of the Small ($Z\approx{0.003}$) and Large Magellanic Clouds ($Z\approx{0.008}$), whereas the \citetalias{Sabhahit2023} model predicts that PISNe only take place at metallicities lower than the one of the Small Magellanic Cloud.

At high redshift, the main contribution to the PISN rate 
comes from very metal-poor stars ($Z<0.002$), for which the mass-loss rates adopted in the \citetalias{Sabhahit2023} and  \citetalias{Chen2015} models behave almost in the same way (Fig.~\ref{fig:pisn_new}). 
However, the (P)PISN rates obtained with models \citetalias{Sabhahit2023} and \citetalias{Chen2015} show some differences even at high redshift. Specifically, at redshift $z\ge{}8$, the rate of the \citetalias{Sabhahit2023} model becomes slightly higher than the rate of the \citetalias{Chen2015} models. 
This happens because of other differences between the \textsc{mesa} and \textsc{parsec} stellar tracks that we already discussed in Section~\ref{sec:threshold}. Specifically,  Fig.~\ref{fig:He_core_mesa_parsec} shows that the final He core mass of the \textsc{parsec} tracks (\citetalias{Chen2015}) at  metallicity $0.0001\leq{}Z\leq{}0.0006$ tends to be lower than the one of the \textsc{mesa} tracks (\citealt{Sabhahit2023}), reducing the overall PISN rate in the \citetalias{Chen2015} models. 

\subsection{The impact of metallicity evolution}
Figure~\ref{fig:pisn_new} shows the contribution of various metallicities to the (P)PISN rate for our two models. Below redshift $z =5$, the two models behave in a completely different way: if we assume the winds by \cite{Sabhahit2023} the main contribution to PISNe (PPISNe) comes from metallicity $Z=0.001-0.002$ ($Z=0.002$), whereas if we adopt the winds by \cite{Chen2015} the dominant metallicity is $Z=0.01$ for both PISNe and PPISNe. At $z>5$, the two models evolve in a more similar way and also lower metallicities become important.

\begin{figure*}
    \centering
    \includegraphics[width=.9\textwidth]{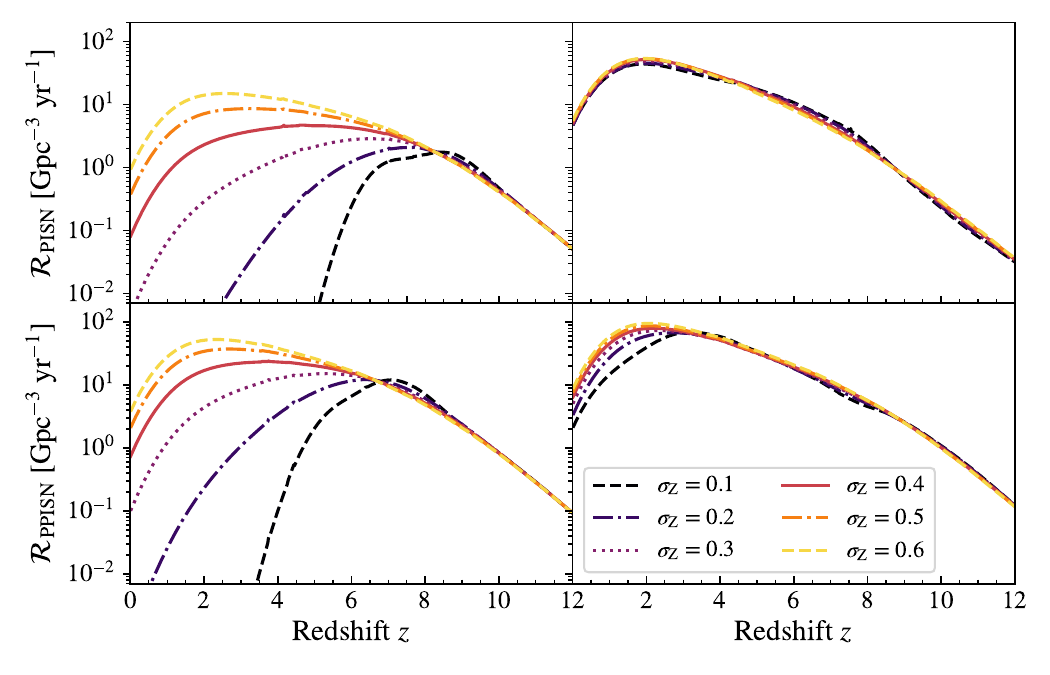}
    \caption{{PISN (upper row) and PPISN (lower row) as a function of the redshift, varying the metallicity spread from $\sigma_{\mathrm{Z}} = 0.1$ ({black} dashed line) to $\sigma_{\mathrm{Z}} = 0.6$ (yellow dashed line). Left-hand (right-hand) panels: models with stellar winds from \protect\citetalias{Sabhahit2023} (from \protect\citetalias{Chen2015}).}
    }
    \label{fig:pisn_sigma}
\end{figure*}

Figure~\ref{fig:pisn_new} assumes our fiducial metallicity spread $\sigma_{\mathrm{Z}}=0.4$ (see Eq. \ref{eq:log_Z}). In Figure \ref{fig:pisn_sigma}, we show the impact of the parameter   $\sigma_{\mathrm{Z}}$ on our results. Specifically, we change the metallicity spread $\sigma_{\mathrm{Z}}$ between 0.1 and 0.6. {Our fiducial value of $\sigma_{\mathrm{Z}}=0.4$ is compatible with observational calibrations \citep[see, e.g.,][]{Chruslinska2019,Santoliquido2021}. However, in Figure~\ref{fig:pisn_sigma} we treat $\sigma_{\rm Z}$ as  a parameter, given the large uncertainties and observational biases \citep[see, e.g.,][for a discussion]{Sgalletta2024}.}

In the results based on models \citetalias{Chen2015}, the rates are mainly influenced by medium-to-high metallicities with a peak at $z \sim 2$ determined by metallicity $Z = 0.01$ (see also Fig. \ref{fig:pisn_new}). As a consequence, changing $\sigma_{\mathrm{Z}}$ has little effect, introducing only minor fluctuations at redshift $z = 7$ for (P)PISN rate density, and leading to PPISN rate densities at $z = 0$ that range from 2 to 8.5 $\text{yr}^{-1}\text{ Gpc}^{-3}$.  
This occurs because of the high metallicity threshold for PISN in \citetalias{Chen2015} models. In this case, the rate density is dominated by the peak of the SFRD, with the spread of the metallicity playing only a minor role \citep{Gabrielli2024}. As a result, the rate is almost insensitive to $\sigma_\text{Z}$.

In contrast, when enhanced winds are considered, there is no contribution from metallicities $Z>2 \times 10^{-3}$ for PISNe ($Z>4 \times 10^{-3}$ for PPISNe). This leads to a strong dependence on metallicity evolution, allowing the PISN rate to be always below the upper limit estimated by \cite{Schulze2024}, even when $\sigma_\mathrm{Z}\sim{0.5}$.
The PISN rate density at redshift $z = 0$ even falls below the lower limit set by \cite{Schulze2024} when assuming $\sigma_{\mathrm{Z}}<0.3$. In this case, a different value of $\sigma_{\mathrm{Z}}$ shifts also the expected peak from $z \sim 7$ (assuming $\sigma_{\mathrm{Z}} = 0.1$) to $z \sim 2$ (assuming $\sigma_{\mathrm{Z}} = 0.6$). These results depend on our simplifying assumption that the distribution of metallicities at a given redshift follows a log-normal distribution around the mean value. We will study more detailed distributions in a follow-up study. Model \citetalias{Chen2015} is almost unaffected by the spread $\sigma{}_{\rm Z}$ because the threshold metallicity for (P)PISNe is so high that there is always a subpopulation of VMSs that end their life with a (P)PISN, even at low redshift. The rate becomes then more sensitive to the IMF (maximum stellar mass and slope of the mass function) than to the metallicity spread.

{\cite{Gabrielli2024} found a similar behavior. They modeled the metallicity as a log-normal distribution centered on the fundamental metallicity relation (FMR; \citealp{Curti2020}), with a dispersion $\tilde{\sigma}_{\mathrm{Z}}$. When adopting the metallicity threshold $Z_{\text{th}} = 0.017$, varying $\tilde{\sigma}_{\text{Z}}$ between 0.15 and 0.70 leads to nearly identical PISN rates at redshift $z = 0$. In contrast, for a lower threshold ($Z_{\text{th}} = 0.0015$), the same variation in $\tilde{\sigma}_{\text{Z}}$ results in a discrepancy of about five orders of magnitude (see Tab. \ref{tab4}). This indicates that the PISN rate is determined not only by $\sigma_{\text{Z}}$, but by the interplay between $\sigma_{\text{Z}}$ and the metallicity threshold $Z_{\text{th}}$.}

\begin{table}[!ht]
\begin{center}
\begin{tabular}{l|c|c}
\toprule
    Model & $\mathcal{R}_{\protect\mathrm{PISN},\,z=0}$ & $\mathcal{R}_\protect\mathrm{PPISN,\,{}z=0}$\\
     & $[\mathrm{yr}^{-1}\,\mathrm{Gyr}^{-3}]$ & $[\mathrm{yr}^{-1}\,\mathrm{Gyr}^{-3}]$\\
    \hline
    This work (\citetalias{Sabhahit2023} winds) & $8\times10^{-2}$ & 0.8\\

    This work (\citetalias{Chen2015} winds) & 5.8 & 6.5\\
    
    \citetalias{Briel2022} (empirical) & 6.6 \\
    
    \citetalias{Briel2022} (\textsc{eagle}) & 0.8 \\
    
    \citetalias{Briel2022} (\textsc{IllustrisTNG}) & 1.3\\
    
    \citetalias{Briel2022} (\textsc{Millennium}) & 5.1 \\
    
    \citetalias{Tanikawa2023} (fiducial) & 1.7 \\
    
    \citetalias{Tanikawa2023} ($-3\sigma$) & 0.02\\
    
    \citetalias{Gabrielli2024} ($\tilde{\sigma}_Z = 0.15$, $Z_{\text{th}} = 0.0015$) & $3\times10^{-4}$\\
    
    \citetalias{Gabrielli2024} ($\tilde{\sigma}_Z = 0.35$, $Z_{\text{th}} = 0.0015$) & 1.2\\
    
    \citetalias{Gabrielli2024} ($\tilde{\sigma}_Z = 0.70$, $Z_{\text{th}} = 0.0015$) & 22\\

    \citetalias{Gabrielli2024} ($Z_{\text{th}} = 0.017$) & 1970\\
    
\bottomrule    
\end{tabular}
\end{center}
\caption{PISN and PPISN rate density at redshift $z = 0$ from \citetalias{Briel2022}, \citetalias{Tanikawa2023}, and \citetalias{Gabrielli2024}. 
For \citetalias{Briel2022}, models Empirical, \textsc{eagle}, \textsc{IllustrisTNG} and \textsc{Millennium} refer to different models for the star formation rate density evolution. For \citetalias{Tanikawa2023}, models fiducial and $-3\sigma$ assume different rates of the $^{12}$C($\alpha$,$\gamma$)$^{16}$O reaction. The different values reported for \citetalias{Gabrielli2024} assume three different values of the metallicity dispersion $\tilde{\sigma}_{\mathrm{Z}}$ and two different metallicity threshold for PISN $Z_{\text{th}}$.} 
\label{tab4}
\end{table}

\section{Discussion}\label{sec:discussion}


The  stellar wind model proposed by \citetalias{Sabhahit2023} has a strong impact on the evolution of VMSs. 
The high mass-loss rate in the optically thick regime causes the most massive stars at metallicity $Z\ge{}0.004$ to end their life with a He core mass lower than 32~M$_{\odot}$, without entering the (pulsational) pair-instability regime. In contrast, at metallicity $Z = 0.001$, even stars with masses between 250 and 500 M$_{\odot}$ can undergo PISNe. 
This has important consequences on the expected mass function of black holes (Torniamenti et al., in prep.). 
Here, we have found that the enhanced mass-loss rate lowers the maximum metallicity for PISNe  from $Z \sim 1.7\times10^{-2}$ to $Z \sim 2\times10^{-3}$, as also shown by \citetalias{Sabhahit2023}. The absence of contributions by higher metallicities suppresses the predicted peak  of the PISN rate at redshift $z\sim{2}$ and reduces the expected rate at redshift $z = 0$ by two orders of magnitude compared to the \textsc{parsec} stellar tracks \citepalias{Chen2015}. The PISN rate density of our \textsc{mesa} models falls within the observational  constraints estimated by \citet{Schulze2024}: $0.009 \text{ Gpc}^{-3} \mathrm{ yr}^{-1} \leq \mathcal{R}_{\mathrm{PISN}} \leq 0.7 \text{ Gpc}^{-3} \mathrm{ yr}^{-1}$.


To better comprehend the critical impact of stellar winds and metallicity on the (P)PISN rate, we compare our results with previous works. 
{A summary of all PISN rate densities at $z = 0$ discussed here is provided in Table \ref{tab4}.} 
\citetalias{Briel2022} demonstrated that the variation of the star formation rate model widely affects the PISN rate. Their empirical model aligns with our results from model \citetalias{Chen2015}, showing a peak around $z \sim 2$ and a similar value at $z = 0$ of approximately $\mathcal{R}_{\mathrm{PISN}} \sim 6 \text{ yr}^{-1}\,\mathrm{Gyr}^{-3}$. In the other models they considered the PISN rate peak is at $z \sim{ 3}$, resulting in a lower rate at $z = 0$, but still exceeding the more recent observational constraints.

\citetalias{Tanikawa2023} explored the impact of  variations in the $^{12}$C($\alpha$,$\gamma$)$^{16}$O reaction rate, predicting a peak of the PISN rate around $z \sim 6$. This higher redshift may be attributed to the different metallicity distribution they assumed. As already discussed by \citetalias{Gabrielli2024}, binary interactions are not expected to significantly impact the predicted rate. Moreover, the inclusion of enhanced stellar winds can lead to wider binaries due to the large amount of mass ejected, reducing the effects of binary evolution.


\citetalias{Gabrielli2024} also investigated how metallicity evolution influences the expected PISN rate. They assumed the metallicity distributed as a log-normal distribution around the fundamental metallicity relation (FMR; \citealp{Curti2020}) with a metallicity spread $\tilde{\sigma}_{\mathrm{Z}}$. Changing $\tilde{\sigma} _{\mathrm{Z}}$ from 0.15 to 0.7 and the maximum metallicity to have PISN $Z_{\text{th}}$ from $1.5\times10^{-3}$ to $1.7\times10^{-2}$, they observed a change of several orders of magnitude in the PISN rate, largely exceeding the upper limits by \cite{Schulze2024} when $\tilde{\sigma}_\mathrm{Z}\gtrsim{}0.35$ and $Z_{\text{th}} \geq 1.5\times10^{-3}$. In our work, the metallicity spread $\sigma_\mathrm{Z}$ is defined as the standard deviation of a log-normal distribution around the average metallicity evolution of stars (Eq.~\ref{eq:log_Z}). Hence, it is not directly comparable with the parameter $\tilde{\sigma}_{\mathrm{Z}}$ adopted by \citetalias{Gabrielli2024}, even if both parameter quantify the dispersion of the stellar metallicity distribution (see the discussion by \citealt{Sgalletta2024}). 

As shown in the left-hand panels of Fig. \ref{fig:pisn_sigma}, the PISN rates of our \textsc{mesa} models are also very sensitive to  metallicity, but, considering $\sigma_\mathrm{Z} \leq 0.5$, the results are consistent with the observational upper limit. We do not expect that a more detailed modeling of metallicity evolution would affect our results. Indeed, since the \citetalias{Sabhahit2023} models do not include contributions from high metallicities, the high-metallicity tail predicted by the log-normal distribution of $Z$ has no impact on the outcome. 

{Overall, previous studies able to reproduce the observed PISN rate (see Tab. \ref{tab4}) either adopt a value that falls in the tail of the observational distribution for some of the parameters, or use a metal-dependent SFRD derived from cosmological simulations. For instance, \citetalias{Tanikawa2023} adopt a value of the $^{12}{\rm C}(\alpha{},\gamma)^{16}{\rm O}$ reaction rate that is 3$\sigma$ lower than the measured one, whereas 
\citetalias{Briel2022} assume the metal-dependent SFRD derived from the \textsc{eagle} cosmological simulation, which generally yields a higher median stellar metallicity than empirical models \citep[e.g.,][]{Artale2020}. Here, we show that our approach, based on a stellar wind model calibrated against observational data \citepalias{Sabhahit2023} and data-driven prescription for the SFRD \citep{Harikane2022},  results in a PISN rate density in agreement with the recent observational constraints \citep{Schulze2024} without assuming extreme values for nuclear reaction rates or other parameters. Moreover, due to optically thick winds, the \citetalias{Sabhahit2023} model yields a metallicity threshold ($Z\sim$ 0.003) above which the PISN is completely avoided, regardless of the maximum stellar mass considered.}

{In a companion paper \citep{Boco2025}, the authors argue that the \citetalias{Sabhahit2023} model, coupled with fast rotation or higher overshooting parameter, can also explain the existence of companion-less WR stars in the Small Magellanic Cloud \citep{Schootemeijer2021}. Additional tests and comparisons are needed, to address whether such models are compatible with other observational benchmarks, e.g. the Humphreys-Davidson limit (\citealt{Humphreys1979}).}


\section{Summary} \label{sec:summary}

Here, we have studied the impact of the new stellar wind model by \citetalias{Sabhahit2023} on the rate of pair-instability supernovae (PISNe) and pulsational pair-instability supernovae (PPISNe), by integrating new very massive star (VMS) tracks with \textsc{mesa}  \citep{Paxton2019} and by convolving our results with a metal-dependent star formation history \citep{Santoliquido2021}.

The model by \citetalias{Sabhahit2023} describes the transition from optically thin to optically thick winds  resulting in high mass-loss rates ($\dot{M}\gtrsim{}10^{-4}$ M$_\odot$) already during the main sequence of VMSs, for metallicities $Z\gtrsim{}0.0004$. This quenches the growth of VMS' cores (Fig.~\ref{fig:He_core_mesa_parsec}) and prevents the onset of PISNe at metallicity $Z\ge{}2\times{}10^{-3}$, even if we assume a maximum ZAMS mass as high as 500~M$_\odot$ (\citetalias{Sabhahit2023}).

We find that the PISN (PPISN) rate density at $z=0$ is $\mathcal{R}_\mathrm{PISN}=0.08$ Gpc$^{-3}$ yr$^{-1}$ ($\mathcal{R}_\mathrm{PPISN}=0.8$ Gpc$^{-3}$ yr$^{-1}$) for the \citetalias{Sabhahit2023} wind model, if we assume a metallicity dispersion $\sigma_\mathrm{Z}=0.4$ (Fig.~\ref{fig:pisn_new} and Table~\ref{tab4}). By changing the metallicity dispersion $\sigma_\mathrm{Z}$ from 0.1 to 0.6, the local (P)PISN rate density changes by several orders of magnitude (Fig.~\ref{fig:pisn_sigma}), because it is extremely sensitive to the fraction of star formation that occurs above the metallicity threshold for (P)PISNe.

For $\sigma{}_\mathrm{Z}=0.4-0.5$, our expected local PISN rate density ($\mathcal{R}_\mathrm{PISN,\,z=0}=0.08-0.4$ Gpc$^{-3}$ yr$^{-1}$) is consistent with the observational constraints ($0.009<\mathcal{R}_\mathrm{PISN}/$Gpc$^{-3}$ yr$^{-1}<0.7$) inferred by \cite{Schulze2024}. Notably, this result does not require to cap the maximum stellar mass or to assume an unrealistically thin metallicity dispersion.

In our \citetalias{Sabhahit2023} models, the PISN rate density grows up to a few $\times{}$~Gpc$^{-3}$~yr$^{-1}$ at redshift $z=2-6$. The main contribution to PISNe and PPISNe comes from stars with metallicity $Z=0.001-0.002$.

We have compared our new \textsc{mesa} models with the \textsc{parsec} stellar tracks reported by \cite{Costa2025}, adopting  the fitting formulas by \citet{Vink2001} for O-type stars, with corrections for electron scattering introduced by \citetalias{Chen2015}. The \textsc{parsec} tracks lead to a higher maximum metallicity ($Z\ge{}1.7\times{}10^{-2}$) for PISNe if we assume the same maximum ZAMS mass. As a consequence, the PISN rate density at $z=0$ is much higher ($\sim{}6$ Gpc$^{-3}$ yr$^{-1}$), above the observational upper limit.

Our results confirm the importance of stellar winds to describe the evolution of VMSs, the rate of (P)PISNe, and the expected mass function of black holes (Torniamenti et al., in prep.).

\begin{acknowledgements}
 We wholeheartedly thank Gautham Sabhahit and collaborators for sharing their \textsc{mesa} inlists (\url{https://github.com/Apophis-1/VMS_Paper1}, \url{https://github.com/Apophis-1/VMS_Paper2}). Our work would not have been possible without this invaluable open-science input.
We thank {Guglielmo Costa}, Marco Dall'Amico, Varsha Ramachandran, Andreas Sander, Raffaele Scala, Jorick Vink{, Gautham Sabhahit, Daniel Pauli and Amedeo Romagnolo} for useful discussions.   MM, GI, and ST acknowledge financial support
from the European Research Council for the ERC Consolidator grant DEMOBLACK, under contract no. 770017. MM and ST also acknowledge financial support from the German Excellence Strategy via the Heidelberg Cluster of Excellence (EXC 2181 - 390900948) STRUCTURES.  
ST acknowledges financial support from the Alexander von Humboldt Foundation for the Humboldt Research Fellowship.
GI  acknowledges financial support from the La Caixa Foundation for the La Caixa Junior Leader fellowship 2024.
GI also acknowledges financial support under the National Recovery and Resilience Plan (NRRP), Mission 4, Component 2, Investment 1.4, - Call for tender No. 3138 of 18/12/2021 of Italian Ministry of University and Research funded by the European Union – NextGenerationEU.
The authors acknowledge support by the state of Baden-W\"urttemberg through bwHPC and the German Research Foundation (DFG) through grants INST 35/1597-1 FUGG and INST 35/1503-1 FUGG.
We use the \textsc{mesa} software (\url{https://docs.mesastar.org/en/latest/}) version r12115; \citep{Paxton2011, Paxton2013, Paxton2015, Paxton2018, Paxton2019}. We used \textsc{sevn} (\url{https://gitlab.com/sevncodes/sevn}) to generate our BBHs catalogs \citep{Spera2019,Mapelli2020,Iorio2023}. We used \textsc{trackcruncher} (\url{https://gitlab.com/sevncodes/trackcruncher}) \citep{Iorio2023} to produce the tables needed for the interpolation in \textsc{sevn}. We used \textsc{cosmorate} (\url{https://gitlab.com/Filippo.santoliquido/cosmo_rate_public}) \citep{Santoliquido2020,Santoliquido2021} to compute the rate densities.
    This research made use of \textsc{NumPy} \citep{Harris20}, \textsc{SciPy} \citep{SciPy2020}, \textsc{Pandas} \citep{Pandas2024}. For the plots we used \textsc{Matplotlib} \citep{Hunter2007}.

\end{acknowledgements}

\bibliographystyle{aa}
\bibliography{ref}

\appendix
\section{\textsc{mesa} models}\label{app}

\begin{figure}
    \centering
    \includegraphics[width=.5\textwidth]{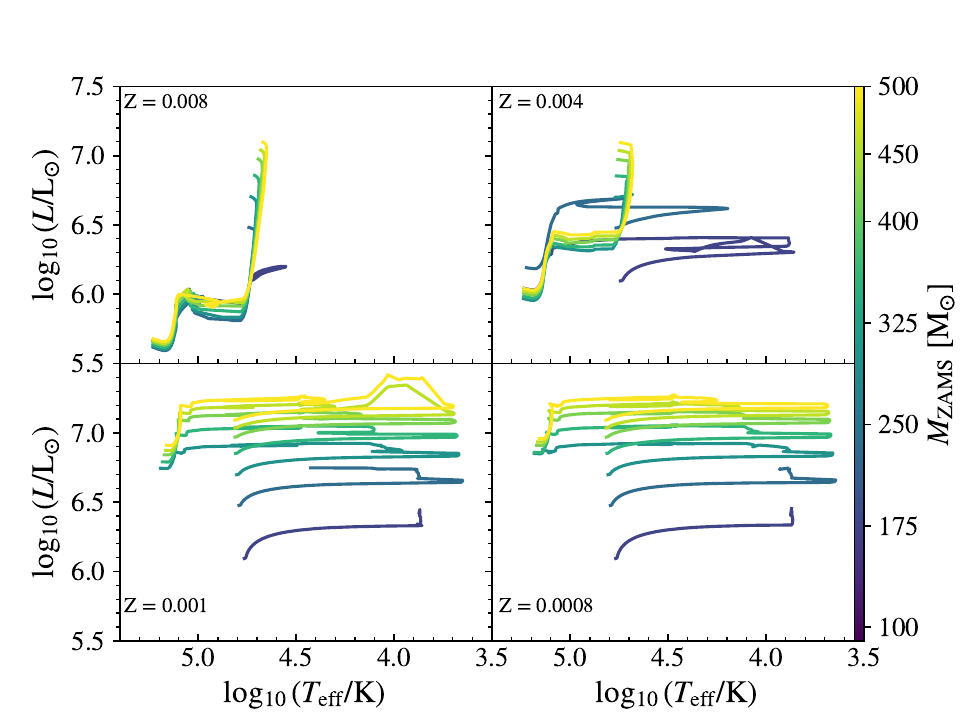}
    \caption{HR diagram of the simulated models. We report four selected metallicities: $Z = 0.008$, 0.004, 0.001, and 0.0008.}
    \label{fig:HR}
\end{figure}

\begin{figure}
    \centering
    \includegraphics[width=.5\textwidth]{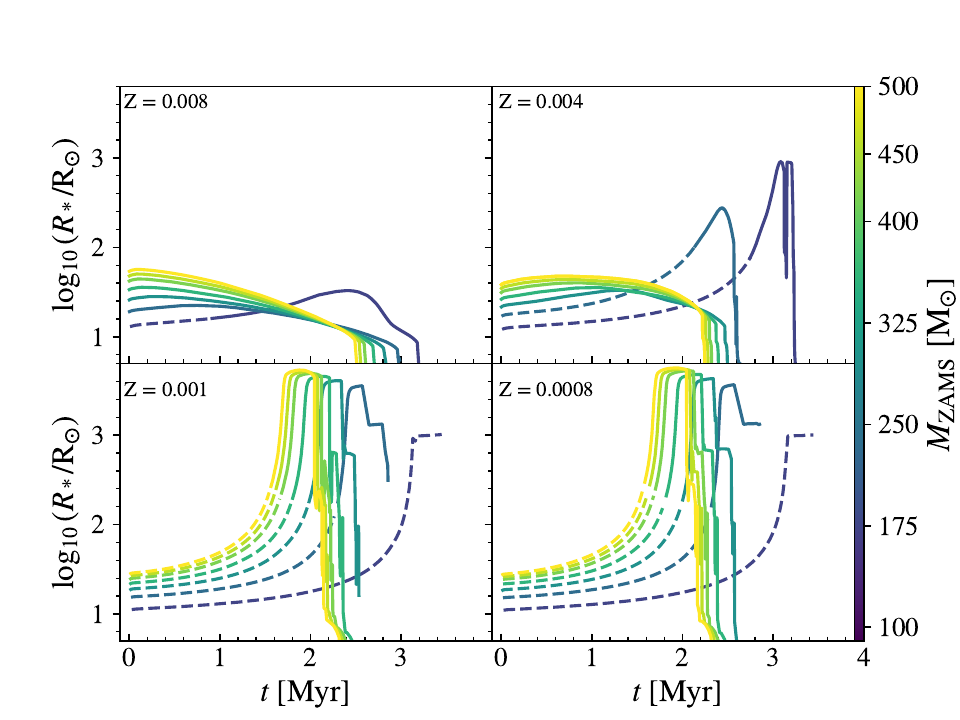}
    \caption{Time evolution of the radius. We report four selected metallicities: $Z = 0.008$, 0.004, 0.001, and 0.0008. The solid lines show the evolutionary phase in which the stars enter the optically thick regime ($\Gamma_{\mathrm{Edd}} > \Gamma_{\mathrm{Edd, tr}}$), while the dashed lines represent the optically thin phase of stellar winds.}
    \label{fig:radius}
\end{figure}

\begin{figure}[h!]
    \centering
    \includegraphics[width=.5\textwidth]{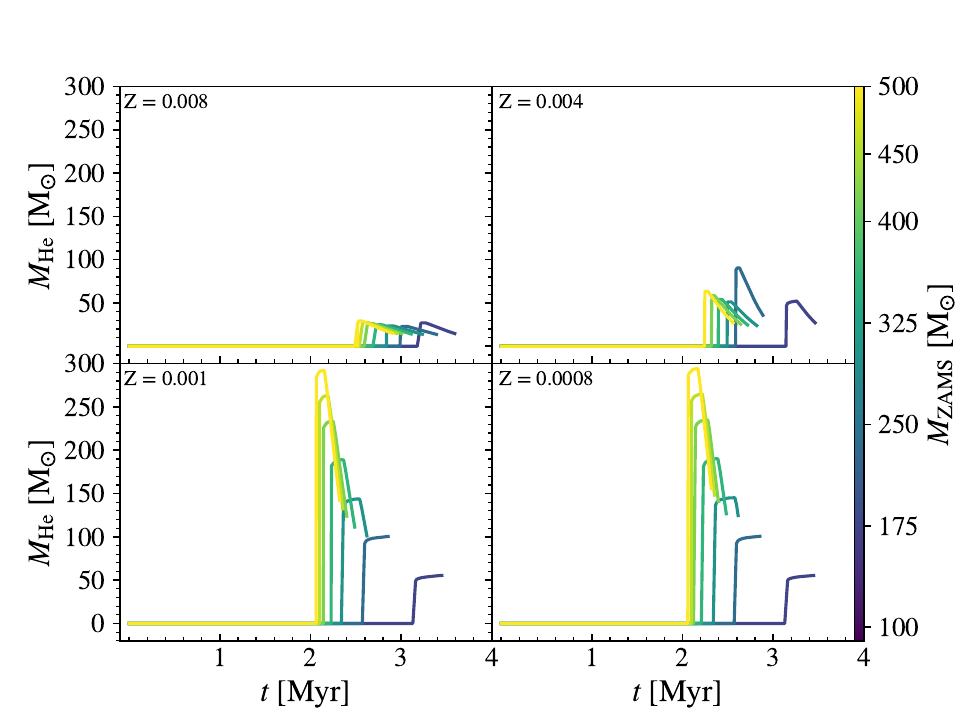}
    \caption{He core mass evolution as a function of time. From top-left to bottom-right: $Z = 0.008$, 0.004, 0.001, and 0.0008.}
    \label{fig:He_mass}
\end{figure}

Here, we summarize the main properties of our \textsc{mesa} models, including 
the evolution of luminosity ($L$) and effective temperature ($T_\mathrm{eff}$) in the Hertzsprung-Russell (HR) diagram (Fig. \ref{fig:HR}), the radius  (Fig. \ref{fig:radius}) and the He core mass (Fig. \ref{fig:He_mass})  until the end of core-He  burning. 

Figure \ref{fig:HR} shows the tracks in the HR diagram. All the models at low metallicity ($Z\leq{}0.001$) do not develop optically thick winds during the main sequence. 
These models evolve horizontally in the HR diagram towards cooler temperatures and the mass loss is not high enough to cause a drop in luminosity. Certain models at higher metallicity, e.g. the $M_{\mathrm{ZAMS}} = 100 \,{}\mathrm{ M}_{\odot}$ model at $Z = 0.008$, go back to the blue part of the HR diagram  after entering the high-$\Gamma_{\mathrm{Edd}}$ phase. At these metallicities, the more massive stars enter the high-$\Gamma_{\mathrm{Edd}}$ regime just after the ZAMS, delineating the boundary between two distinct evolutionary behaviors: these stars evolve vertically in the HR diagram, exhibiting a steep decline in luminosity while remaining within a narrow range of temperature.

The radii of stars that enter the enhanced wind regime immediately after the ZAMS rapidly decrease as a consequence of the large amount of mass loss: indeed, in these cases, the H-rich envelope is quickly lost and also mass from the inner regions of the star is expelled (Fig. \ref{fig:radius}). In the models that switch to high-$\Gamma_{\mathrm{Edd}}$ regime during the main-sequence, the radius reaches the peak at $\sim 2 \text{ Myr}$ and then decreases. When the models enter this regime in the very last part of the main sequence, the mass loss is not able to contrast the inflation and the radius grows up to $\log(R/\mathrm{R}_{\odot}) \sim 3.5$ in a short timescale. In all the models, the radii decrease rapidly after the ignition of the He core.

Figure \ref{fig:He_mass} shows the evolution of the He-core mass, while Table \ref{tab3} reports the final values of the He core mass, at the end of core-He burning. At $Z = 0.008$, the mass of the He core  decreases quickly because of wind mass loss, and all models end their life with $13.7 \text{ M}_{\odot} \leq M_{\mathrm{He, f}} \leq 20 \text{ M}_{\odot}$. These systems cannot enter the pair-instability regime and undergo a core-collapse supernova. Lowering the metallicity, at $Z = 0.004$, only stars with $150 \text{ M}_{\odot} \leq M_{\text{ZAMS}} \leq 200 \text{ M}_{\odot}$ can experience PPISNe, while in all the other cases we find $M_{\mathrm{He, f}} < 32 \text{ M}_{\odot}$. At $Z = 0.001$, stars can experience different fates: only the $M_{\mathrm{ZAMS}} = 50 \text{ M}_{\odot}$ model undergoes a core-collapse supernova; models with $M_{\mathrm{ZAMS}} = 75\,,\,100 \text{ M}_{\odot}$  go through PPISN, while models with $125 \text{ M}_{\odot} \leq M_{\mathrm{ZAMS}} \leq 450 \text{ M}_{\odot}$ can experience PISNe. Finally, if $M_{\mathrm{ZAMS}} = 475\,,\,500 \text{ M}_{\odot}$ the model directly collapses onto a black hole. At even lower metallicity ($Z = 0.0008$), the initial stellar mass range of systems that can undergo a PISN is reduced: $125 \text{ M}_{\odot} \leq M_{\mathrm{ZAMS}} \leq 350 \text{ M}_{\odot}$.\\
\begin{table}
\begin{center}
    \begin{tabular}{c|c|c|c|c}
    & \multicolumn{4}{|c}{Final He core mass [M$_{\odot}$]}\\
    \cline{2-5}
    \hline
    $M_{\mathrm{ZAMS}}$ [M$_{\odot}$]& $Z$ = 0.008 & $Z$ = 0.004 & $Z$ = 0.001 & $Z$ = 0.0008\\
    \hline
    50 & 19.9 & 25.3 & 24.7 & 24.7 \\
    75 & 17.7 & 32.4 & 39.9 & 40.1 \\
    100 & 14.7 & 27.2 & 55.5 & 55.4 \\
    125 & 13.8 & 28.9 & 70.6 & 70.6 \\
    150 &  13.7 & 32.2 & 85.6 & 85.6 \\
    175 & 13.7 & 35.7 & 100.5 & 100.7 \\
    200 & 13.8 & 39.4 & 109.7 & 115.6 \\
    225 & 13.8 & 23.8 & 105.4 & 129.2 \\
    250 & 13.9 & 23.9 & 101.1 & 124.5 \\
    275 & 13.9 & 24.1 & 104.6 & 124.1 \\
    300 & 14.1 & 24.4 & 107.3 & 127.2 \\
    325 & 14.2 & 24.6 & 111.3 & 126.6 \\
    350 & 14.3 & 24.9 & 116.1 & 131.7 \\
    375 & 14.4 & 25.2 & 118.7 & 132.9 \\
    400 & 14.5 & 25.5 & 123.8 & 140.9 \\
    425 & 14.7 & 25.9 & 127.7 & 142.2 \\
    450 & 14.8 & 26.2 & 132.2 & 148.1 \\
    475 & 14.9 & 26.6 & 135.8 & 151.6 \\
    500 & 15.1 & 27.1 & 142.1 & 155.5 \\
    \hline
    \end{tabular} 
\end{center}
\caption{Final mass of the He core of the single stellar model at $Z = 0.008$ (second column), $Z = 0.004$ (third column), $Z = 0.001$ (fourth column) $Z = 0.0008$ (fifth column). In the first column is reported the initial total mass of the corresponding star.}
\label{tab3}
\end{table}

\section{\textsc{parsec} models}\label{app:parsec}

Our comparison sample is based on \textsc{parsec} stellar tracks, which have been extensively discussed by \citet{Costa2025}; see also \citet{Bressan2012, Chen2015, Costa2022, Nguyen2022, Iorio2023, Costa2023} for additional references. Here, we briefly summarize their main properties.

For the winds of massive O-type stars, \textsc{parsec} adopts the fitting formulas by \citet{Vink2000} and \cite{Vink2001}, with a correction introduced by \citet{Chen2015} to account for the effects of electron scattering, based on the models of \citet{Graefener2008}. In this framework, the mass-loss rate of an O-type star scales as $\dot{M} \propto Z^{\beta}$, where $Z$ is the metallicity and the exponent $\beta$ depends on the Eddington ratio $\Gamma_{\rm Edd}$:
$\beta = 0.85$ for $\Gamma_{\rm Edd} < 2/3$,
$\beta = 2.45 - 2.4\times\Gamma_{\rm Edd}$ for $2/3 \le \Gamma_{\rm Edd} < 1$,
and $\beta = 0.05$ for $\Gamma_{\rm Edd} \ge 1$ \citep{Chen2015}.
This formalism provides a fit to the models by \citet{Graefener2008} and \citet{Vink2011}, incorporating the dependence of mass loss on the Eddington ratio.

For Wolf–Rayet (WR) stars, \textsc{parsec} adopts the prescriptions by \citet{Sander2019}, which successfully reproduce the observed properties of Galactic WR stars of type C (WC) and type O (WO).
 

\textsc{parsec} models include internal mixing by adopting the mixing-length theory \citep[MLT;][]{Bohm-Vitense1958}, calibrated on the Sun with a mixing-length parameter $\alpha_\mathrm{MLT} = 1.74$ \citep{Bressan2012}. Convective regions are identified using the Schwarzschild criterion \citep{Schwarzschild1958}, as implemented in \citet{Bressan1981}.
The \textsc{parsec} tracks used in this work assume an overshooting parameter of $\lambda_\mathrm{ov} = 0.5$, where $\lambda_\mathrm{ov}$ represents the mean free path of convective elements beyond the convective boundary, expressed in units of the pressure scale height. In addition, an envelope undershooting distance of $\Lambda_{\text{env}} = 0.7$ pressure scale heights is adopted.


\end{document}